# Influence of density and release properties of UC$_x$ targets on the fission product yields at ALTO

Julien Guillot[a], Brigitte Roussière[a], Sandrine Tusseau-Nenez[b], Denis S. Grebenkov[b], Maxime Ignacio[b]

[a] Laboratoire de physique des 2 infinis Irène Joliot-Curie, CNRS/IN2P3 UMR 9012 - Université Paris Saclay, F-91406 Orsay Cedex

[b] Laboratoire de Physique de la Matière Condensée, CNRS, Ecole Polytechnique UMR 7643 - IP Paris - Route de Saclay, F-91128 Palaiseau Cedex

*Corresponding author: guillotjulien@ipno.in2p3.fr, tel: +33(0)169157160

________________________________________________________________________________

________________________________________________________________________________

**Abstract**:

To study the influence of the structural properties of UC$_x$ targets on their release properties, several types of targets using different precursors (carbon and uranium) were synthesized, characterized, irradiated and heated leading to the determination of the released fractions of eight elements. In this article, the production rates of these targets are estimated under the use conditions at ALTO, i.e. with targets bombarded by an electron beam (10 μA, 50 MeV). We have simulated the fission number produced using the FLUKA code. Then, we have determined the release efficiency as a function of the half-life of the isotopes using average diffusion coefficients deduced for the elements studied previously. Finally, we compare the production rates obtained from the various targets and conclude that the target must be adapted to the element studied. It is crucial to find in each case the best compromise between the target density and the release efficiency.

________________________________________________________________________________

1. Introduction:

Advances in nuclear physics have led to a growing demand for the production of intense exotic beams. The intensity of a radioactive beam ($I$) is given by equation (1) and depends on the intensity of the primary beam ($I_p$), the number of target nuclei exposed to the primary beam ($N$), the production cross section ($\sigma$) of the nucleus of interest as well as the release ($\varepsilon_{release}$), ionization ($\varepsilon_{ionization}$) and transport ($\varepsilon_{transport}$) efficiencies:

$$I = I_p \times N \times \sigma \times \varepsilon_{release} \times \varepsilon_{ionization} \times \varepsilon_{transport} \qquad (1)$$

Then producing an intense radioactive beam implies to optimize each of these parameters, a problem involving different fields of physics and technique and particularly difficult to solve because possible interactions between the various contributions can worsen the overall efficiency. So, maximizing $I$ requires to optimize jointly:

- $I_p$, that depends on the performance of the accelerator
- $\sigma$, that is obtained by nuclear physics measurements and depends on the energy of the collision and the nature of the collision partners
- $N$, that, to be large enough, needs the use of thick targets; the targets must be stable at high temperature and radiation-resistant, both properties depending on physicochemical properties of materials
- $\varepsilon_{release}$, which involves diffusion in the material, desorption on its surface and effusion by porosities of the target, phenomena studied by chemistry and solid state physics
- $\varepsilon_{ionization}$, which depends on the element studied and involves ion-source technology
- $\varepsilon_{transport}$, which is an ionic-optics problem.

Some of these parameters are defined by the installation used: this is the case in particular of the incident beam. At ALTO we use an electron beam of 50 MeV energy and 10 μA intensity to induce photofission [1].

In the last two decades, many laboratories have focused on improving the performance of the target ion source system (TISS) and in particular the release efficiency of the material used for the ISOL target design [2]. In recent years, at the ALTO facility, we have conducted studies to understand and determine the factors that improve the release efficiency of ISOL targets. A systematic study on synthesis parameters (milling and mixing of precursor powders, pressing, carburation) led to the development of nanostructured-target synthesis protocols[3]–[8]. Fourteen different sample types were developed. Two types of uranium powders were used: a uranium oxide ground powder ($UO_2$) and a uranium oxalate hydrate powder (OXA). Three different carbon sources were also used: graphite, carbon nanotubes (CNT) and graphene. Three mixing methods have been used. For the "conventional protocol" (CP), the two precursor powders of uranium and carbon were mixed with the help of an automatic mixer. For the "developing protocol" (DP), the uranium powder and the different carbon sources were dispersed separately in isopropanol under ultrasound and then mixed with C/U ratios equal to 5, 6 or 7 using a hand blender. For the "graphene protocol" (GP), the graphene was obtained by exfoliation of the graphite in a blender in presence of a sodium dodecyl sulphate solution. The previously dispersed uranium oxide powder is then added in the blender (see appendix A Table 1 and ref [7]). After having characterized their physicochemical properties (see appendix A, Table 2), the samples were irradiated with a deuteron beam delivered by the tandem accelerator of the laboratory. The released fractions (RFs) of 11 elements were measured by gamma spectrometry [7]. Analysis of the results by principal component analysis has allowed us to highlight strong correlations between release properties and certain structural characteristics, notably high open porosity distributed on small-size pores. In order to extract, from the measured RFs, average diffusion coefficients and obtain analytical expressions for the release efficiency, the diffusion model developed by Fujioka et al. [9] and Kirchner [10] for three target geometries (foil, fiber, sphere) has been extended to the geometry of our samples, namely a cylinder.

In this paper, we investigate the interaction between the in-target production and the release properties by comparing, for radioisotopes differing by their chemical nature and half-lives, the yields expected to be released by the fourteen above-mentioned targets presenting different microstructures. First, as the in-target production is directly correlated to the number of fissions produced in the target, the number of fissions induced by the Bremsstrahlung radiation emitted by the 50 MeV energy electrons impinging the target were calculated using the FLUKA code [11], [12]. Then the number of nuclei released by various targets was determined, as a function of the isotope half-life, using the release properties measured previously.

2.  Benchmarch of FLUKA for photofission

As a first step, we sought to determine whether the results obtained by the FLUKA code [11], [12] for photofission were in agreement with the estimates made by Diamond [13] and Oganessian *et al.* [14]. Indeed, these estimates were used in the development of the ALTO project to assess that, with a 50 MeV and 10 μA electron beam, about $10^{11}$ fissions per second would be produced in a uranium carbide target.

All the calculations presented below were performed using the FLUKA 2011.2x-6 version and the 2.3-0 version of the Flair interface.

The first calculations were made by bombarding with electron beams of different energies (10, 25, 30, 45, 50 and 70 MeV) a tungsten converter placed in front of a $^{238}U$ metal target (ρ = 18.95 g.cm$^{-3}$). Both are cylinders of 2 cm diameter and 0.2 cm thickness for the tungsten converter and 2 cm thickness for the uranium target. The beam is assumed to have a Gaussian profile in the two directions perpendicular to its propagation axis, with a full width at half maximum (FWHM) equal to 0.235 cm in both directions. This configuration corresponds to the experimental conditions described in the work of Oganessian *et al.* [14]. For each energy, $10^8$ events were simulated.

Figure 1a shows the spectra of photons obtained. The fission rates were calculated in two ways in order to test the respective roles played by the two quantities involved in the determination of the fission rate, namely the photon flux and the fission cross section. In the first case, the photon spectra obtained from the FLUKA calculations (figure 1a) were convoluted with the photofission reaction cross section resulting from the compilation of A.V. Varlamov *et al.* [15]. In the second case, the fission rate was directly deduced from the isotopic distribution of the reaction products calculated by FLUKA. The results obtained by these two approaches are displayed in figure 1b and compared to those presented by Oganessian et al.[14].

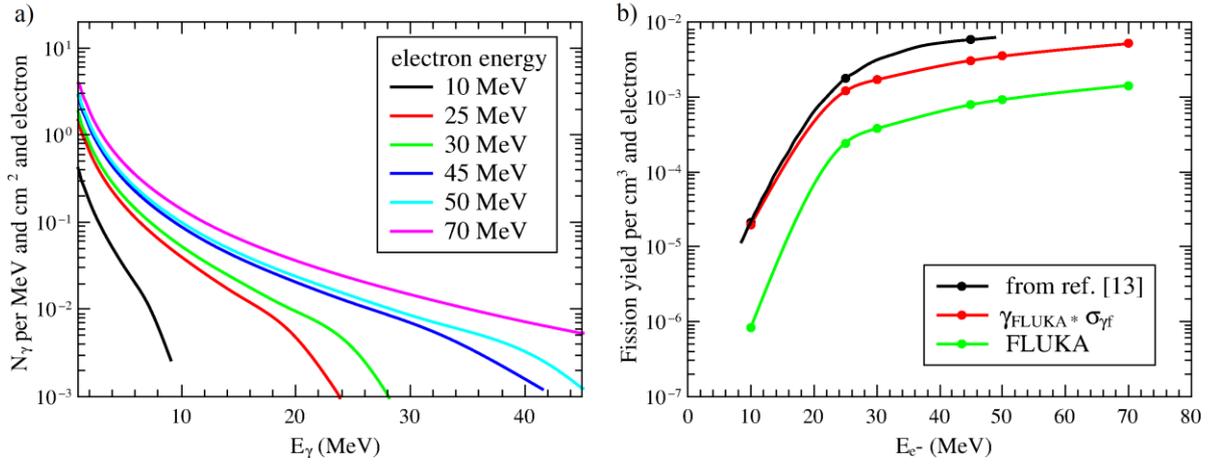

Figure 1: a) Photon spectra calculated for a $^{238}$U target ($\rho$ = 18.95 g.cm$^{-3}$) and electron beams of different energies b) Fission rates obtained in a $^{238}$U target as a function of the electron beam energy ($\gamma_{FLUKA}$ is taken from figure 1a and $\sigma_{\gamma f}$ taken from [14]).

The fission rates calculated in the frame of the first approach (red curve) are in excellent agreement with the results published in ref. [14] for electron energies up to 25 MeV, but they are lower by a factor 1.9 for higher electron energy. This 1.9 factor can be attributed to a slight underestimate of the photon flux by FLUKA. The fission rates obtained by the second approach (green curve) are clearly underestimated, by a factor 7.5 for electron energies greater than 20 MeV. This reduction factor can be splitted into two contributions: the first one, equal to 1.9, due to the underestimate of the photon flux and the second one, equal to 3.9, due to the underestimation of the photofission cross section. The underestimation in the photofission description by FLUKA-2011 has already been reported on the FLUKA forum [16].

We can conclude that for the electron energy available at ALTO, 50 MeV, the photofission cross section is underestimated by FLUKA by a factor of 3.8 and the fission rate by a factor of 6.9. As the purpose of our FLUKA simulations is to compare targets with different densities in terms of production, a relative answer is sufficient. However, when considering absolute quantities, this factor 6.9 should be taken into account.

In order to highlight the influence of the converter, a second calculation was performed using the same geometry as before but with 25 and 50 MeV electron beams hitting directly a metal target of $^{238}$U ($\rho$ = 18.95 g.cm$^{-3}$). Figure 2 shows that the photon distributions obtained in both cases are very similar. The fission rates obtained per cm$^3$ and per electron amount respectively to 2.42×10$^{-4}$ and 2.94×10$^{-4}$ with and without converter for a 25 MeV electron beam and to 9.18×10$^{-4}$ and 9.92×10$^{-4}$ with and without converter for a 50 MeV electron beam. The presence of a 2 mm thick tungsten converter reduces the number of fissions, but its influence is much less significant with higher beam energy. As long as the electron beam is stopped in the target, the number of fissions created in the target is higher without converter. This is the case at ALTO where long (19.3 cm) uranium carbide targets ($\rho$ = 3.82 g.cm$^{-3}$) and a 50 MeV electron beam are used. In addition, the intensity of the beam delivered by ALTO (10 µA) leads to a very reasonable beam power (500 W) easily bearable by the target. So it is relevant not to use a converter at ALTO. Therefore in the following, all the simulations are carried out without converter.

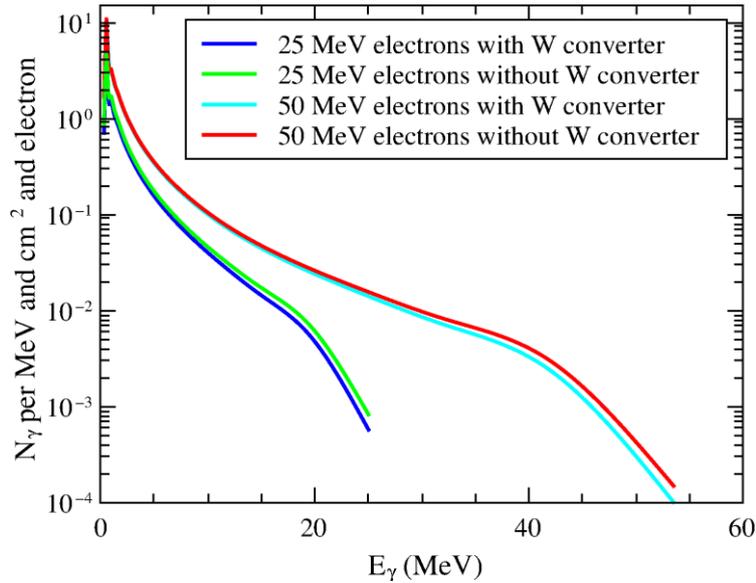

Figure 2: Photon fluence calculated for 25 and 50 MeV electron beams on a $^{238}$U target with or without W converter

3. Calculation of the fission number produced in UC$_x$ targets with various densities

In our previous study [8], we modelled a uranium carbide target as a cylinder of 13 mm in diameter and 19.3 cm in length. These dimensions correspond to the standard dimensions of the targets used in both ISOLDE and ISOL-ALTO installations. The 50 MeV electrons of the ALTO primary beam have a much lower penetrating power than the 1.7 GeV protons used at ISOLDE. Consequently, as shown in ref. [8], almost 90 % of the fissions are created in the first 6 cm of the conventional target (sample No.8 PARRNe BP897 CP). By reducing the length of the target to 6 cm, the number of fissions produced remains almost the same ($1.97 \times 10^{10}$ versus $2.22 \times 10^{10}$, for a 10 µA beam intensity) whereas the target volume and consequently the amount of radioactive waste is divided by a factor 3.2. In addition, with a target reduced in size but centred in the oven, the fission products are created closer to the opening leading to the transfer tube before ionization and in a volume smaller than before, which will improve the effusion efficiency.

In order to calculate the number of fissions produced in targets elaborated from the 14 samples developed in ref. [7], nine simulations performed with $6 \times 10^8$ events have been carried out under conditions very similar to those described in ref. [8]. An UC$_x$ target, defined by its density and length, is located in the middle of a tantalum tube with 20 mm diameter, 200 mm length and 1 mm thickness. Before reaching the target, the beam (Gaussian profile with FWHM = 0.3915 cm in both directions perpendicular to its propagating direction) passes through a 1 mm thick graphite window, a graphite tube (1 mm thick, variable length) and a 1mm thick graphite plug. The length of each target was determined in order to produce in these short targets ~ 90 % of the fissions created in a target with standard size and same density.

Table 1 summarizes the data describing the UC$_x$ samples relevant for the simulations as well as the number of fissions calculated by FLUKA assuming a 10 µA electron beam. The samples Nos. 9, 10, 11 and 12 differ from samples Nos. 8, 1, 2 and 3 only by a 12-day heating treatment at 1800 °C; in our previous studies [7], [8] the sample No. 7 only served as a reference to check reproducibility. Therefore, no specific calculation was performed for these 5 samples.

Table 1. Description of the samples developed in ref. [6], density and length of the target used in the FLUKA simulations and number of fissions calculated for an electron beam of 10 µA intensity and 50 MeV energy.

| No. | Samples | Apparent density (g.cm$^3$) | Length (cm) | Number of fissions | Radioactive-waste reduction ratio (%) |
|---|---|---|---|---|---|
| 1 | UO$_2$ ground + CNT CP | 1.35 | 10.2 | 8.64×10$^9$ | 47.2 |
| 2 | UO$_2$ ground + CNT DP | 1.68 | 9 | 1.04×10$^{10}$ | 53.4 |
| 3 | UO$_2$ ground + graphene GP | 3.76 | 6 | 1.94×10$^{10}$ | 68.9 |
| 4 | OXA + graphite CP | 3.42 | 6.4 | 1.82×10$^{10}$ | 66.8 |
| 5 | OXA ground + CNT DP | 1.35 | 10.2 | 8.64×10$^9$ | 47.2 |
| 6 | OXA + CNT DP | 1.1 | 11.3 | 7.11×10$^9$ | 41.5 |
| 8 | PARRNe BP897 CP | 3.82 | 6 | 1.97×10$^{10}$ | 68.9 |
| 9 | PARRNe BP897 CP 12d | 3.82* | 6 | 1.97×10$^{10}$ | 68.9 |
| 10 | UO$_2$ ground + CNT CP 12d | 1.35* | 10.2 | 8.64×10$^9$ | 47.2 |
| 11 | UO$_2$ ground + CNT DP 12d | 1.68* | 9 | 1.04×10$^{10}$ | 53.4 |
| 12 | UO$_2$ ground + graphene GP 12d | 3.76* | 6 | 1.94×10$^{10}$ | 68.9 |
| 13 | UO$_2$ ground + CNT-5moles DP | 2.78 | 7.1 | 1.64×10$^{10}$ | 63.2 |
| 14 | UO$_2$ ground + CNT-7moles DP | 1.27 | 10.5 | 7.76×10$^9$ | 45.6 |

* density taken equal to that of the 12d non-heated sample

Figure 3 shows the distribution of electrons, gammas produced by Bremsstrahlung, and fissions induced for two short targets: the conventional No. 8 target (ρ = 3.82 g.cm$^{-3}$) and the No. 14 target that exhibited the best released fractions of the fission products studied but that is one of the least dense (ρ = 1.27 g.cm$^{-3}$). These distributions exhibit patterns very similar to those obtained in the first 6 or 10.5 cm of a standard-size target (see figure 9 in ref. [8]).

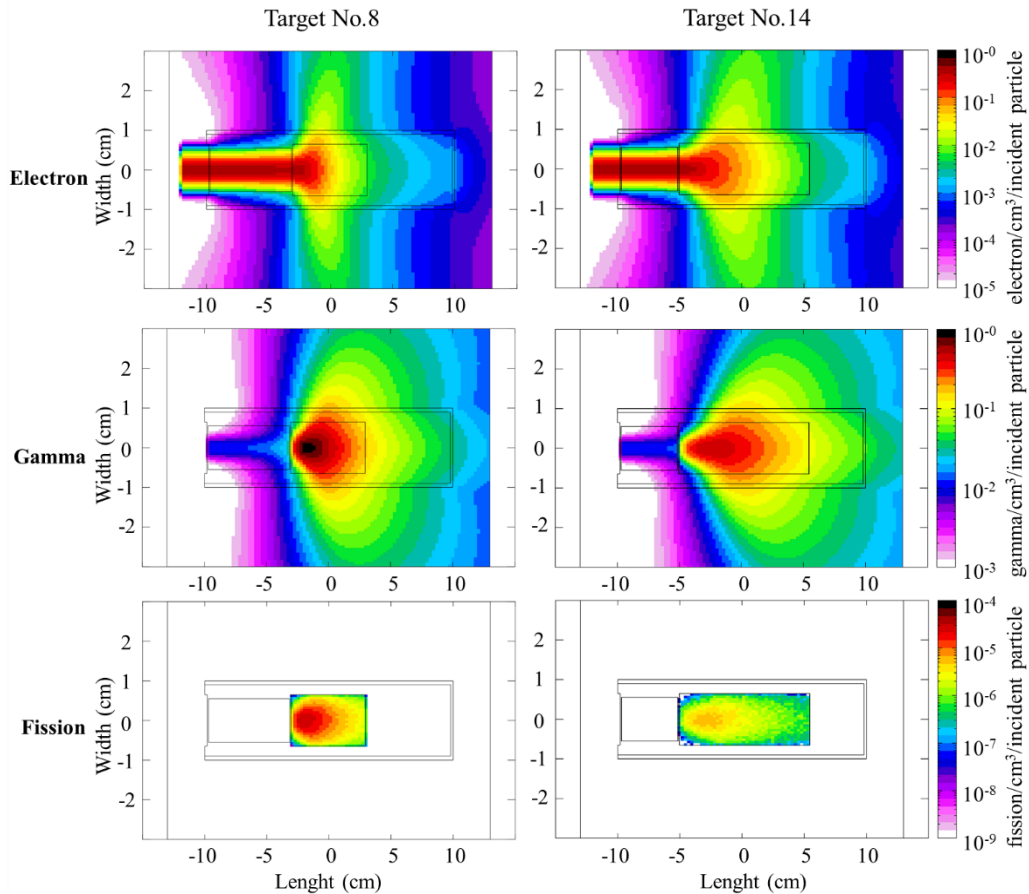

Figure 3: Electron, gamma and fission distributions obtained for No.8 (PARRNe BP897 CP) and No.14 (UO$_2$ ground + CNT – 7 moles DP) targets.

As expected, Table 1 shows that the number of fissions produced in the target is proportional to the density of the sample used and that the lower the $UC_x$ density, the greater the length of the short target. In order to quantify the benefit obtained by using short target, we calculate the reduction ratio in the amount of radioactive waste. It ranges from 41.5 % for the target with the lower density (sample No. 6) to 68.9 % for the higher density target (sample No. 8). Whatever the $UC_x$ sample used, the reduction ratio is high and it is far better to use short target. This will be done now at ALTO, for example the length of a conventional target (sample No. 8) will be 6 cm.

In the following, we will discuss whether a release efficiency enhanced by the target microstructure can compensate the loss in the fission amount due to a low density.

4. Release efficiency for a cylindrical geometry

In our previous study we extended the Fujioka and Arai calculations [9] to a cylindrical geometry to determine the release efficiency at the target exit ($\varepsilon_{RF}$) of a given isotope of an element (see ref. [8] for more details). We used the following equation:

$$\varepsilon_{RF}(\lambda) = \frac{32}{\pi^2} \sum_{k,m=1}^{\infty} \frac{\pi^2 \frac{(2m-1)^2}{L^2} + \frac{J_{0,k}^2}{R^2}}{J_{0,k}^2 (2m-1)^2 \left( \pi^2 \frac{(2m-1)^2}{L^2} + \frac{J_{0,k}^2}{R^2} + \frac{\lambda}{D} \right)} \qquad (2)$$

with $R$ (cm) and $L$ (cm) the radius and thickness of the pellet, $D$ (cm$^2$.s$^{-1}$) the diffusion coefficient (average of diffusion in grain and effusion through porosity), $J_{0,k}$ the kth positive root of the zeroth order Bessel function of the first kind and $\lambda$ (s$^{-1}$) the radioactive decay constant of the isotope.

Equation (2) can be re-written in the following way, which is slightly more convenient from the numerical point of view as it converges faster:

$$\varepsilon_{RF}(\lambda) = 1 - \frac{32}{\pi^2} \sum_{k,m=1}^{\infty} \frac{1}{J_{0,k}^2 (2m-1)^2 \left( 1 + \frac{\pi^2 (2m-1)^2 D}{\lambda L^2} + \frac{J_{0,k}^2 D}{\lambda R^2} \right)} \qquad (3)$$

Moreover, one of the two sums can be computed exactly. For instance, one can get the following expression:

$$\varepsilon_{RF}(\lambda) = 1 - \frac{8}{\pi^2} \sum_{m=1}^{\infty} \frac{1}{(2m-1)^2 \left( 1 + \frac{\pi^2 (2m-1)^2 D}{\lambda L^2} \right)} \left( 1 - \frac{2 I_1(S_m)}{S_m I_0(S_m)} \right) \qquad (4)$$

where $S_m = \sqrt{\lambda R^2/D + \pi^2 (2m-1)^2 R^2/L^2}$ and $I_n(z)$ are the modified Bessel functions of the first kind. The obvious advantage of this formula is that there is only one sum over $m$ instead of two sums over $k$ and $m$. One also gains that there is no need to evaluate the zeros $J_{0,k}$ of the Bessel function.

In the following, all the release efficiencies have been calculated using Equation (4), since on a standard laptop the calculations are completed nearly one hundred times faster with Equation (4) than with Equation (2).

5. Expected production rates from $UC_x$ targets with various densities

Firstly, the average diffusion coefficients $D$ have been extracted from the released fractions measured in ref. [8] using the formula describing the diffusion process in a cylinder (Equation (2) in ref. [8]). Then using these average diffusion coefficients, the released efficiencies $\varepsilon_{RF}$ are calculated according to the procedure described in Section 3. Finally, the production rates $\Phi$ are obtained by using the following formula (5):

$$\Phi = N_{fission} \times Y \times \varepsilon_{RF} \qquad (5)$$

where $N_{fission}$ is the fission number generated in the target determined by FLUKA, $Y$ the fission yield of the considered isotope and $\varepsilon_{RF}$ the release efficiency of this isotope.

Figure 4 shows the production rates released by targets Nos. 1 to 14 relatively to those calculated for the conventional target PARRNe No. 8 as a function of the half-life of the isotope and for the elements studied in ref. [8], namely krypton (Kr), strontium (Sr), tin (Sn), antimony (Sb), tellurium (Te), iodine (I), cesium (Cs) and barium (Ba).

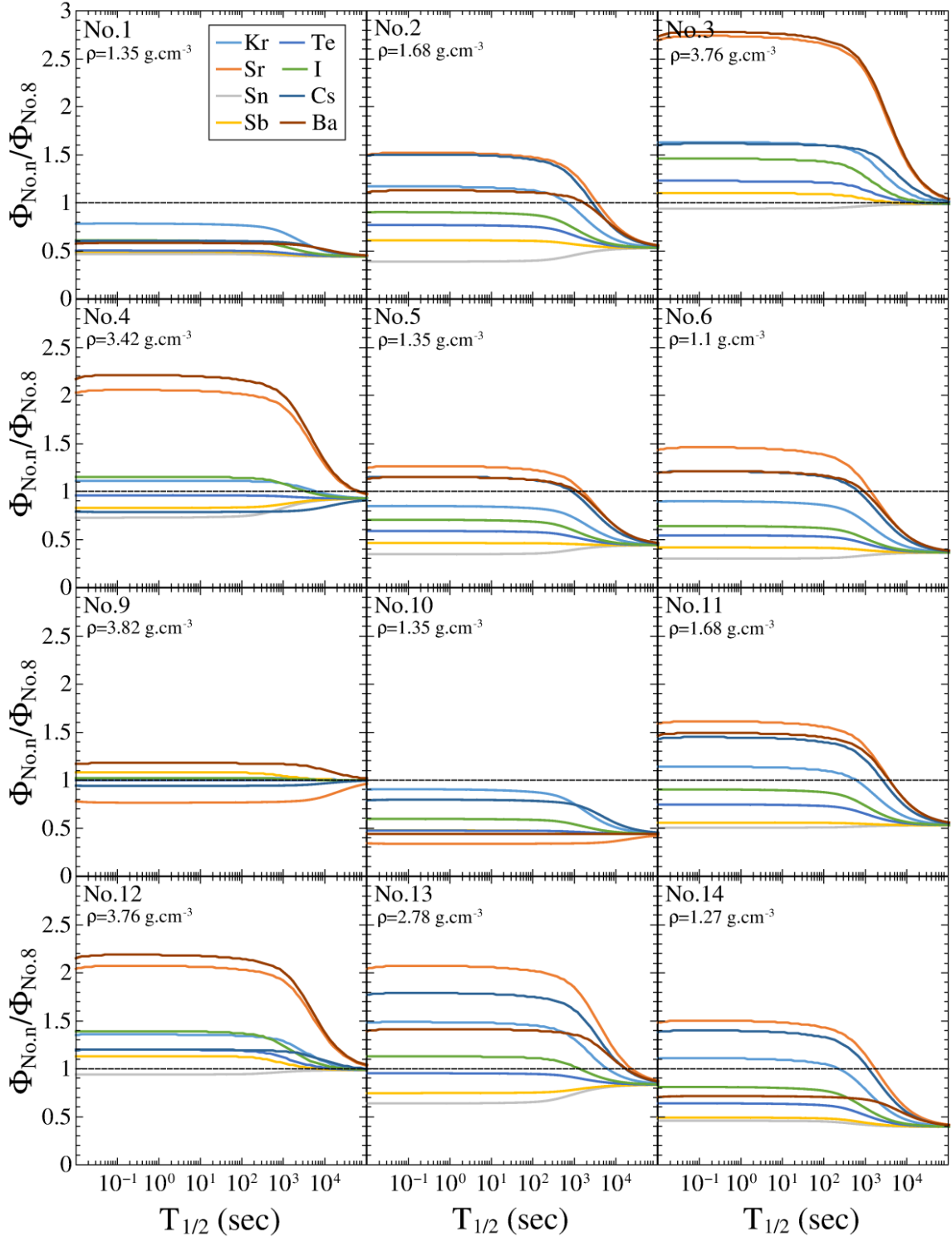

Figure 4: Rates of Kr, Sr, Sn, Sb, Te, I, Cs and Ba released by targets Nos. 1 to 14 relatively to the conventional target PARRNe No. 8 as a function of the half-life of the isotope for a temperature of 1768 °C

For the long-lived isotopes, the ratio of the exit rates between the two targets ($\Phi_{No.n} / \Phi_{No.8}$) tends to the ratio of the number of fissions produced in the two targets ($N_{fission\ No.n} / N_{fission\ No.8}$). The shorter the isotope half-life, the more visible the influence of the release efficiency. And so, for isotopes with half-lives less than 100 seconds, the curves reach a plateau (called in Figure 5 $\Phi_{No.n} / \Phi_{No.8\ asymptotic}$). When the plateau value is greater than 1, the target tested releases the concerned element more efficiently than the conventional No.8 target, which means that better release properties can compensate for production loss due to lower density of the target. However, the interplay between release efficiency and density is rather complex. The $\Phi_{No.n} / \Phi_{No.8}$ value depends strongly on the element and the highest values are obtained for Kr, Ba, Sr and Cs.

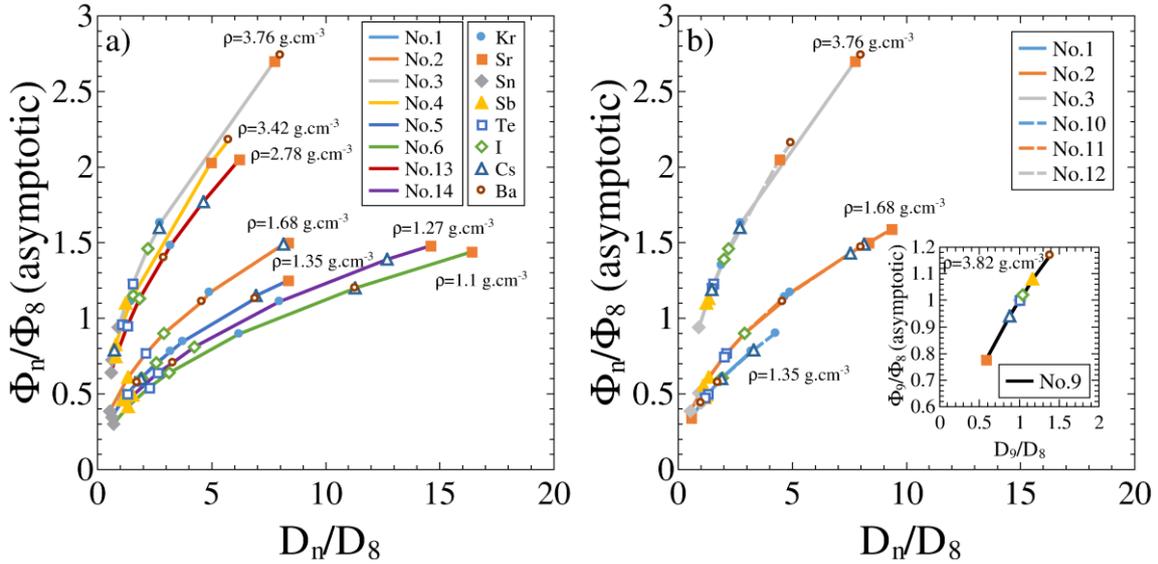

Figure 5: Exit rates of Kr, Sr, Sn, Sb, Te, I, Cs and Ba isotopes with half-life less than 100 seconds from targets Nos. 1 to 14 relatively to the target No. 8 as a function of the ratio of the diffusion coefficient in the target *n* over the diffusion coefficient in the conventional target. a) representation of the samples only carburized b) representation of the carburized samples before and after the heating treatment of 12 days. Data describing the sample No. 9 having weak amplitude, they are drawn in the insert for sake of legibility.

The targets Nos. 1, 5 and 10 have the same density ($\rho$ = 1.35 g.cm-3) but display different behaviours (see Figure 4). Whereas all the $\Phi_{No.n} / \Phi_{No.8}$ values are lower than 1 for targets Nos. 1 and 10, this is not the case for target No. 5 for Sr, Ba and Cs. For these elements, the release properties of sample No. 5 are better than those of samples Nos 1 and 10 [8]. These differences observed in the release are related to the microstructure of the samples. Indeed, it has been shown that the more efficient samples in terms of release have grains and aggregates with small size and a high porosity distributed on small diameter pores [8]. While samples Nos 1 and 10 were both prepared according to the CP protocol leading to large aggregates and a significant contribution of large diameter pores (they differ only by the 12-day heating treatment applied only to sample No. 10), sample No. 5 was obtained using the DP protocol that reduces the aggregate size and leads exclusively to the formation of small diameter pores.

Among the targets with low density, the target No. 6 ($\rho$ = 1.1 g.cm-3) and the target No. 14 ($\rho$ = 1.27 g.cm-3), which is the best sample in terms of physicochemical and release properties [8], outperform target No. 8 for three elements: Sr, Ba and Cs and Kr, Sr and Cs, respectively. Both samples were obtained using CNTs as carbon precursor allowing to obtain a high porosity and were prepared according to the DP protocol which quasi-exclusively leads to small diameter pores as well as grains and aggregates of small sizes, prerequisites for good release properties [8].

Targets Nos 2, 13 and 14 ($\rho$ = 1.68, 2.78 and 1.27 g.cm-3 respectively) were prepared according to the DP protocol using ground $UO_2$ and CNT as uranium and carbon precursors, but with different molar ratio between the carbon and uranium quantities used (C/U = 6, 5 and 7 respectively). These three samples are defined by very similar microstructures in terms of crystallite, grain and aggregate sizes, porosities and pore size distributions ensuring, for all the elements studied, good release properties improved by the carbon excess. Despite their densities lower than that of sample No. 8, the samples Nos 2, 13 and 14 show higher rates after release for 4 (Sr, Cs, Kr and Ba), 5 (Sr, Cs, Kr, Ba and I) and 3 (Sr, Cs and Kr) elements respectively (see figure 4). In addition, sample No. 13 outperforms the other two because its higher density compensates for its release properties reduced by its smaller C/U ratio.

For targets with density higher and close to that of the conventional target No. 8 ($\rho$ = 3.82 g.cm-3), namely targets Nos. 3 ($\rho$ = 3.76 g.cm-3), 4 ($\rho$ = 3.42 g.cm-3), 9 ($\rho$ = 3.82 g.cm-3) and 12 ($\rho$ = 3.76 g.cm-3), the impact of the release properties on the exit rates is easier to observe. For target No. 9, which has the same microstructure as the conventional target with the addition of a 12-day heating treatment, the $\Phi_{No.9} / \Phi_{No.8}$ values are close to 1 whatever the element. This shows the long-term stability at high temperature of the conventional target. Target No. 4 exhibits an improvement of the exit rates for 4 elements (Ba, Sr, I and Kr) directly related to the enhancement in the release properties between targets No. 8 and 4. Both targets were prepared according to the CP protocol but using different uranium precursor, $UO_2$ for sample No. 8 and OXA for sample No. 4. The use of $UO_2$ leads to a pore distribution centered on 3 and 10 μm diameters, whereas, when OXA is used, the pore diameter is mainly around 3 μm. This improves slightly the open porosity and the release of some elements, in particular Ba and Sr [8].

Targets Nos. 3 and 12 outperform, in terms of rates after release, the conventional target for all elements except Sn. The release fractions measured for Sn are very similar for all the samples studied [8]; consequently, the $\Phi_{No.3} / \Phi_{No.8}$ and $\Phi_{No.12} / \Phi_{No.8}$ values for Sn reflect only the slight change in density between the targets Nos. 3 (or 12) and 8. Targets 3 and 12 differ from the conventional one by the carbon precursor used, graphene instead of graphite, and then the protocol used, GP instead of CP. The samples Nos. 3 and 12 have physicochemical properties close to the conventional one but, surprisingly, better release properties. The target No. 3 achieves the best overall rates after release but its microstructure is not fully stable at high temperature. Indeed, the only change in the preparation of sample No. 12 from that of sample No. 3 is an additional 12-day heating treatment which induces an Ostwald ripening resulting in an abnormal growth of the grains that ultimately increases the release time of the fission products of the elements released by diffusion. Compared to sample No. 3, the parameters describing the microstructure of sample No. 12 are very similar (see appendix A supplementary data) however its microstructure seems to be less homogeneous (see figure 3 of ref. [7]) and exhibits locally large aggregates.

Figure 5 provides an overall view of the results. The $\Phi_{No.n} / \Phi_{No.8}$ values for isotopes with half-life less than 100 seconds are displayed as a function of $D_n / D_8$, the ratio of the diffusion coefficient in the target n over the diffusion coefficient in the conventional target. For each target, the data corresponding to the different elements draw a very smooth curved line with a slope increasing as the density. Looking at Figure 5a anti-clockwise, the targets are displayed by density increasing order: firstly target No. 6, then targets Nos. 14, 1, 5, 2, 13 and finally targets Nos. 4 and 3. The curves corresponding to samples Nos. 1 and 5 are superimposed because both targets have the same density. In the same way, targets Nos. 10, 11 and 12 differ from targets Nos. 1, 2 and 3 respectively, only by a long-term heating at 1800 °C; therefore, having the same density, they overlap, and only three curves appear in Figure 5b. The stability of the microstructure at high temperature can be read directly in Figure 5b: the more stable the microstructure, the nearer the points describing a given element for the heated and non-heated samples.

Figure 5 indicates that samples Nos. 6 and 14 release some specific elements very quickly. If these targets could be densified while keeping the same release properties, they would enable the study of some very short-lived isotopes.

6. Conclusion

In this paper, we studied the interaction between the in-target production and the release properties of the target. Our previous studies had investigated the impact of the microstructure of various $UC_x$ samples on the release properties: the best samples in terms of release had a microstructure composed of small $UC_x$ grains and presenting a high open porosity distributed homogeneously on small pores. However, the samples sharing these characteristics were also the less dense in uranium. The question of how the release properties would compensate the lower density was worth consideration. Thus the yields expected to be released by targets presenting different microstructures and density have been estimated for isotopes differing by their chemical nature and half-lives.

FLUKA simulations have been performed in order to estimate the number of fissions induced in uranium carbide targets with different densities. The target length has been chosen to be well suited for use at ALTO with an electron beam of 50 MeV, i.e. leading to a good compromise between a maximal number of fissions and a minimal radioactive waste volume. It has been shown that, for all the samples studied, the target length necessary to obtain 90 % of the fission number produced in a standard-size target reduces the amount of radioactive waste from ∼ 40 up to 70 % depending of the target density. Then, average diffusion coefficients for eight elements in $UC_x$ targets with different microstructure have been extracted from the release fractions published recently [8], and the impact of the release properties on the rates available out of the targets has been investigated.

The most promising target appears to be the one based on graphene samples. Indeed, this is the only target that gives higher exit rates than those achieved with the conventional target for all the elements studied. Its density is very similar to that of the conventional target, so the higher exit rates obtained are due to its better release properties. However, the latter worsen after a long-term treatment at high temperature. A short-term goal should be to improve the graphene-target synthesis protocol in order to stabilize its microstructure at high temperature. Once this is achieved, this graphene target could replace the PARRNe target as a general-purpose target.

Targets based on carbon nanotubes and prepared according to the so-called "developing protocol" in ref. [8] exhibit exit rates greater than those obtained by the conventional one for some elements among Kr, Sr, Cs, Ba and I. Most of them are less dense in uranium than the PARRNe target, but the lower in-target productions can be compensated by a higher release efficiency. A longer-term goal should be to increase the density in uranium of these carbon-nanotube targets, while maintaining their main structural properties (small grains, high porosity composed of small pores) leading to good release properties. In this way, targets dedicated to specific elements could be developed and allow the study of isotopes with very short half-lives.